\documentclass[]{pasj01}
\draft

\begin{document} 
\Received{}
\Accepted{}

\title{Search for a Non-equilibrium Plasma in the Merging Galaxy Cluster Abell 754}

\author{Shota \textsc{Inoue}\altaffilmark{1,2}%
}
\altaffiltext{1}{Department of Earth and Space Science, Graduate School of Science, Osaka University, 1-1 Machikaneyama-cho, Toyonaka, Osaka 560-0043, Japan}
\altaffiltext{2}{Project Research Center for Fundamental Sciences, Graduate School of Science, Osaka University, 1-1 Machikaneyama-cho, Toyonaka, Osaka 560-0043, Japan}
\email{shota@ess.sci.osaka-u.ac.jp}

\author{Kiyoshi \textsc{Hayashida}\altaffilmark{1,2}}

\author{Shutaro \textsc{Ueda}\altaffilmark{3}}
\altaffiltext{3}{Institute of Space and Astronautical Science (ISAS), Japan Aerospace Exploration Agency (JAXA),
3-1-1 Yoshinodai, Chuo-ku, Sagamihara, Kanagawa 229-8510,
Japan}

\author{Ryo \textsc{Nagino}\altaffilmark{1,2}}

\author{Hiroshi \textsc{Tsunemi}\altaffilmark{1,2}}

\author{Katsuji \textsc{Koyama}\altaffilmark{1,2,4}}
\altaffiltext{4}{Department of Physics, Graduate School of Science, Kyoto University, Kitashirakawa, Oiwake-cho, Sakyo-ku, Kyoto 606-8502, Japan}


\maketitle

\begin{abstract}
Abell 754 is a galaxy cluster in which an ongoing merger is evident on the plane of the sky, from the southeast to the northwest. 
We study the spatial variation of the X-ray spectra observed with Suzaku along the merging direction, centering on the Fe Ly\,$\alpha$ / Fe He\,$\alpha$ line ratio to search for possible deviation from ionization equilibrium. 
Fitting with a single temperature collisional non-equilibrium plasma model shows that the electron temperature increases from the southeast to the northwest. 
The ionization parameter is consistent with that in equilibrium ($n_{\rm e}t>10^{13}$\,s\,cm$^{-3}$) except for a specific region with the highest temperature ($kT=13.3_{-1.1}^{+1.4}$\,keV) where $n_{\rm e}t=10^{11.6_{-1.7}^{+0.6}}$\,s\,cm$^{-3}$. The elapsed time from the plasma heating estimated from the ionization parameter is 0.36--76 Myr at the 90\% confidence level. 
This time scale is quite short but consistent with the traveling time of a shock to pass through that region.
We thus interpret that the non-equilibrium ionization plasma in Abell 754 observed is a remnant of the shock heating in the merger process. 
We, however, note that the X-ray spectrum of the specific region where the non-equilibrium is found can also be fitted with a collisional ionization plasma model with two temperatures, low $kT=4.2^{+4.2}_{-1.5}$\,keV and very high $kT>19.3$\,keV. The very high temperature component is alternatively fitted with a power law model. Either of these spectral models is interpreted as a consequence of the ongoing merger process as in the case of that with the non-equilibrium ionization plasma.
\end{abstract}


\section{Introduction}

Galaxy clusters are the largest virialized structures in the universe, hosting hot plasmas with temperatures of $10^{7-8}$\,K as the intracluster medium (ICM) and dark matters. 
The galaxy clusters have evolved primarily via their subcluster mergers, which are the most energetic phenomenon with largest size shock waves after the big bang. 
Observations show many cases of merging clusters (e.g. 1E0657-56: \cite{Markevitch_2002}, Abell 2256: \cite{Tamura_2011}, CIZA J2242.8+5301: \cite{Ogrean_2014}), and discontinuities in temperature, electron density, or pressure distributions are identified as evidence of shocks in some clusters (e.g. Abell 520: \cite{Markevitch_2005}, Abell 3667: \cite{Akamatsu_2012}). 

When a shock passes through a plasma, the plasma is heated by the shock efficiently. 
The timescale for the thermal equilibration between heated electrons through their Coulomb scatterings is expressed as
\begin{eqnarray}
 t_{\rm e-e} \sim 4\times10^4\,{\rm yr}\,\left(\frac{n_{\rm e}}{10^{-3}\,{\rm cm}^{-3}}\right)^{-1}\left(\frac{T_{\rm e}}{5\times 10^7\,{\rm K}}\right)^{3/2}
\end{eqnarray}
(see \cite{Spitzer_1962}), while that for the collisional ionization equilibrium (CIE) is expressed as
\begin{equation}
 t_{\rm ion}\sim 3\times10^7\,{\rm yr}\,\left(\frac{n_{\rm e}}{10^{-3}\,{\rm cm}^{-3}}\right)^{-1}
\end{equation}
(for Fe He\,$\alpha$ and Fe Ly\,$\alpha$ in a plasma with the temperature $kT$ of several keV; see also \cite{Masai_1984}, \cite{Akahori_Yoshikawa_2010}), where $n_{\rm e}$ and $T_{\rm e}$ are the number density and temperature of electrons, respectively. Since $t_{\rm ion}$ is much longer than $t_{\rm e-e}$, the non-equilibrium ionization (NEI) state can persist for a certain period after the shock passes through the plasma.

Although observational evidence of NEI plasma is widely confirmed for supernova remnants (e.g. \cite{Tsunemi_1986}, \cite{Vink_2003}, \cite{Katsuda_2008}), the ionization state of the ICM is usually taken to be CIE, because $t_{\rm e-e}$ and $t_{\rm ion}$ are much shorter than the evolutional timescale of clusters. 
There are a few works to examine this assumption (e.g. \cite{Fujita_2008}), observational evidence of NEI plasmas in galaxy clusters has not been found yet. 
Nevertheless, in the case of merging clusters, the NEI state might remain specific regions near shocks where the period after the plasma heating is short. 
In fact, numerical simulations of merging clusters (e.g. \cite{Takizawa_1999}, \cite{Akahori_Yoshikawa_2010}) predict the NEI state in some locations in their simulated merging clusters. 
The N -body/SPH simulation carried out by \citet{Akahori_Yoshikawa_2010} shows that the line intensity ratio of Fe ions, Fe Ly\,$\alpha$ / Fe He\,$\alpha$, deviates from that in the CIE state at the corresponding location of the shock. 
If we observe an NEI plasma with the ionization parameter $n_{\rm e}t < 10^{13}$\,s\,cm$^{-3}$, we can quantitatively estimate the timescale after the shock heating.

As mentioned above, the ionization state of the ICM has been examined in a limited number of papers. 
\citet{Fujita_2008} reported the ionization state of the Ophiuchus cluster observed with the Suzaku satellite (\cite{Mitsuda_2007}). 
The Ophiuchus cluster is hot and bright enough for us to examine the line ratio of Fe Ly\,$\alpha$ to Fe He\,$\alpha$ in its spatially resolved X-ray spectra. 
Nevertheless, \citet{Fujita_2008} concluded that the ICM of the Ophiuchus cluster reached ionization equilibrium over the whole cluster. 
\citet{Nevalainen_2010} also examined the Fe line ratios in 6 nearby relaxed clusters observed with the XMM-Newton EPIC primarily for a calibration purpose. 
As a result, temperatures derived from Fe line ratios were consistent with those estimated from the X-ray continua for all the clusters. 
They confirmed not only their calibration accuracy, but also little deviation from the CIE state. 
The ionization state in the ICM has been tested in merging clusters using various plasma emission models (e.g. Abell 2146: \cite{Russell_2012}, CIZA J2242.8+5301: \cite{Akamatsu_2015}). 
Nevertheless, only lower limits of ionization parameters were obtained. 
They found no significant evidence of NEI plasmas even in merging clusters.

In this paper, we search for an NEI plasma in the merging galaxy cluster Abell 754 using the Suzaku data. 
Deviation from the ionization equilibrium would be key information on the past of the merging process. 
We adopt the Hubble constant $H_0=$70\,km\,s$^{-1}$\,Mpc$^{-1}$, $\Omega _m=0.2$ and $\Omega _\lambda=0.8$. Unless otherwise noted, errors correspond to 90\% confidence intervals.

\section{Abell 754}
\label{sec:abell754}

Abell 754 is a well-studied major merger cluster on the plane of the sky at $z = 0.0542$ (\cite{Struble_Rood_1999}). Its surface brightness is elongated along the southeast-northwest direction. 
Previous works studied its galaxy distribution (e.g. \cite{Fabricant_1986}, \cite{Zabludoff_Zaritsky_1995}), the profile of the radio emission (e.g. \cite{Kassim_2001}, \cite{Bacchi_2003}, \cite{Kale_2009}, \cite{Macario_2011}), the X-ray morphology and the temperature structure in the ICM (e.g. \cite{Markevitch_2003}, \cite{Henry_2004}, \cite{Macario_2011}). These works reveal the followings:
\begin{itemize}
\item Neither of the two major clumps in the galaxy distribution corresponds to the X-ray brightness peaks.
\item Spatial variations of the temperature are found with highest temperature $kT>10$\,keV.
\item The 1.4 GHz radio diffuse emission consists of two large components (\cite{Bacchi_2003}), roughly correspond to the two large main galaxy concentrations.
\end{itemize}
These features indicate Abell 754 is undergoing a recent major merger. 
Hydrodynamic simulations carried out by \citet{Roettiger_1998} reproduce the observed X-ray features by considering an off-axis merger between two subclusters with a mass ratio of $2.5:1$. 
They also imply the merger is a very recent one ($<$0.3Gyr). 
\citet{Markevitch_2003} suggests that a more complex picture for the merger in which the third subcluster or a cloud of a cool gas is sloshing independently from its former host subcluster. 
\citet{Macario_2011} first confirmed the shock front with the Mach number of ${\cal M}=1.57^{+0.16}_{-0.12}$ at the southeast part of the cluster by measuring the temperature jump with Chandra. 
They also noticed the edge of one radio emission coincide with this temperature jump. 
In this paper, we call this ''shock-SE''. 
Shock-SE supports the idea that the merger takes place along the southeast and northwest axis, although parameters for the merger are loosely constrained. 

Besides, using the BeppoSAX data, \citet{Fusco-Femiano_2003} reported the $\sim3\sigma$ detection of a hard X-ray excess above $\sim45$\,keV. 
They discuss two possibilities of this excess: the emission from the BL Lac object, 26W20, in the field of view of BeppoSAX, or the inverse Compton emission from relativistic electrons.



\section{Observation and data reduction}

The Suzaku observation of Abell 754 was performed with an effective exposure of 109\,ks from 2007 May 29 to 2007 June 1. 
Suzaku has four X-ray Imaging Spectrometer (XIS; \cite{Koyama_2007}) cameras. 
Three of the XIS (XIS0, 2, 3) and the other (XIS1) employ front-illuminated (FI) CCDs and a back-illuminated (BI) CCD, respectively. 
Since XIS2 has been out of function from 2006 November 9, we use the other three XIS (i.e. XIS0, 1, 3). 
The observation was performed with either the normal $3\times3$ or $5\times5$ mode, and we combined these data in our analysis. 
The non X-ray background (NXB) was produced with the FTOOL {\it xisnxbgen} (\cite{Tawa_2008}) using the night-earth data taken 150 days before and after our observation. 
The NXB was subtracted from the X-ray spectra, while we include the X-ray background in the spectral fitting model. 
The response matrix files and the auxiliary response files (ARFs) were generated by the FTOOL {\it xisrmfgen} and {\it xissimarfgen} (\cite{Ishisaki_2007}), respectively. 
As well as the XIS data, the Hard X-ray Detector (HXD; \cite{Takahashi_2007}, \cite{Kokubun_2007}) data were also employed. 
We use the screened event file of the Si PIN diode (HXD-PIN), including a dead-time correction. 
For the NXB of HXD-PIN, we use the simulated event file supplied by the HXD team based on the {\it LCFITDT} model (\cite{Fukazawa_2009}). 
We apply the response matrix for a point source at the XIS nominal position supplied by the HXD team.



\section{Analysis and result}
\label{sec:analysis_result}

\begin{figure}
 \begin{center}
  \includegraphics[width=12cm]{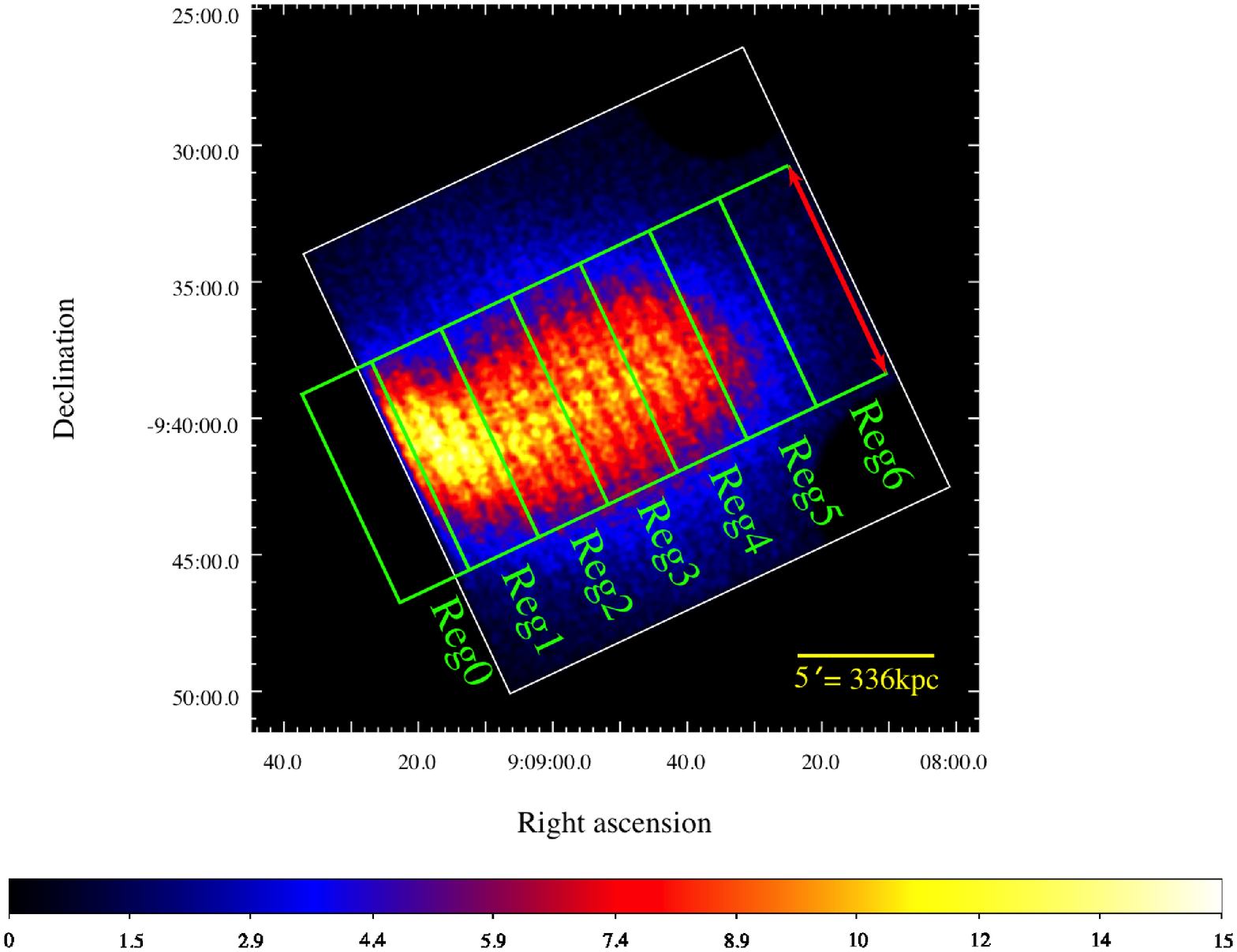} 
 \end{center}
\caption{X-ray image of Abell 754 detected by the XIS0 onboard Suzaku. The image is smoothed with a 2-dimensional gaussian with $\sigma=5\,{\rm pixel}=\timeform{5''.20}$. 
The color scale unit is cts\,Ms$^{-1}$\,pixel$^{-1}$. 
Each green box indicates the analysis region defined in section \ref{sec:analysis_result} and the size of each region is 567\,kpc$\times$189\,kpc. The red arrow indicates the axis to which we project the surface brightness (see section \ref{subsubsec:spatial_distribution}).}\label{fig:A754Image}
\end{figure}

We use XSELECT ver.\,2.4 for extracting images and spectra. 
Figure \ref{fig:A754Image} shows the XIS0 image of Abell 754 in the 2--10\,keV energy band. 
We define six box regions (Reg1--6) along the merging direction. 
We extract the X-ray spectrum from each region. 
In this analysis, we find bright point sources neither in the Suzaku nor XMM-Newton image (\cite{Henry_2004}). 
We employ XSPEC ver.\,12.8.2n to perform spectral fitting. 
X-ray spectra in the 2--10\,keV and in the 2--8\,keV bands are used for the FI CCDs and BI CCD, respectively. 
The FI spectra added together. 
Since \citet{Tamura_2014} reported that the systematic deviation of the energy scales between the FI and BI CCDs is only 0.09\%, we fit FI and BI spectra with the same model parameters. 
Each spectral bin has a minimum count of twenty. 
In these energy bands, the cosmic X-ray background (CXB) is the only X-ray background component we should consider. 
We adopt the powerlaw model with the parameters presented by \citet{Kushino_2002} for the CXB model. 
In this paper, we report the metal abundances relative to the solar values in \citet{Anders_Grevesse_1989}.

In the following analyses, we use the plasma code ver.\,3.0.0 ({\it vvrnei} model\footnote{https://heasarc.gsfc.nasa.gov/xanadu/xspec/manual/XSmodelRnei.html}) installed in XSPEC. 
The {\it vvrnei} model is identical to the Kazik Borkowski's NEI model (\cite{Borkowski_2001}) except that the {\it vvrnei} model has a parameter to specify the initial plasma temperature. 
Unless otherwise noted, we fix this initial temperature at 0\,keV. 
In this paper, we call the {\it vvrnei} model by setting $n_{\rm e}t$ as a free parameter ''NEI model'', while the same model with $n_{\rm e}t$ fixed at $10^{13}$\,s\,cm$^{-3}$ ''CIE model''.


\subsection{Spectral fitting procedures}
\label{subsec:photon_contamination}

The X-ray telescope (XRT) of Suzaku has a half power diameter of $1.8-2.3$ arcmin (\cite{Serlemitsos_2007}). 
Incident photons from one sky region contaminate to other extraction regions. 
In this paper, we fit the spectra extracted from each region simultaneously by taking the photon contamination to each extraction region from its adjacent regions into account. 
We assume the emission of each sky region consists of the position-dependent ICM component (NEI model) and the uniform CXB component (powerlaw). 
The ICM component in each extraction region consists of the ICM emission of the corresponding sky region and those from the adjacent sky regions. 
Proper auxiliary response files (ARF) are prepared using {\it xissimarfgen} (\cite{Ishisaki_2007}). 
In order to estimate the photons from the outer part of the Suzaku field of view, we define Reg0 at the southeastern side of Reg1 (figure \ref{fig:A754Image}). 
On the other hand, contamination to Reg6 from its northwest adjacent region is neglected. 
We input the XMM-Newton images (0.2--12\,keV) of Reg0--3 and the Suzaku images (1--5\,keV) of Reg4--6, as the photon distributions needed for {\it xissimarfgen}. 

First the spectra extracted from each region (Reg1--6) are fitted simultaneously, employing the NEI models for the ICM component. 
In the fitting, the normalization ratio between the emission model of Reg0 and that of Reg1 is fixed at the ratio of the count rates between them estimated by the XMM-Newton data. 
Other parameters of the Reg0 model are linked to those of the Reg1 model. 
The comparison with the X-ray image and the numerical simulation suggests that Abell754 is a merger on the plane of the sky (\cite{Roettiger_1998}). 
We thus assume the redshifts of all the regions are equal and obtain them to be $0.05393_{-0.00018}^{+0.00012}$. 
Note that this redshift is consistent with that determined by optical observations (\cite{Struble_Rood_1999}).
The best fit parameters for the ICM components determined here are employed to model the contamination from the adjacent sky regions. 
In the following analysis, we perform the spectral fitting for each sky region individually, by fixing spectral model of adjacent region.

\subsection{Line intensity ratio of Fe Ly\,$\alpha$ / Fe He\,$\alpha$}
\label{subsec:Fe_intensity_ratio}

\begin{figure}
 \begin{center}
  \includegraphics[width=8cm]{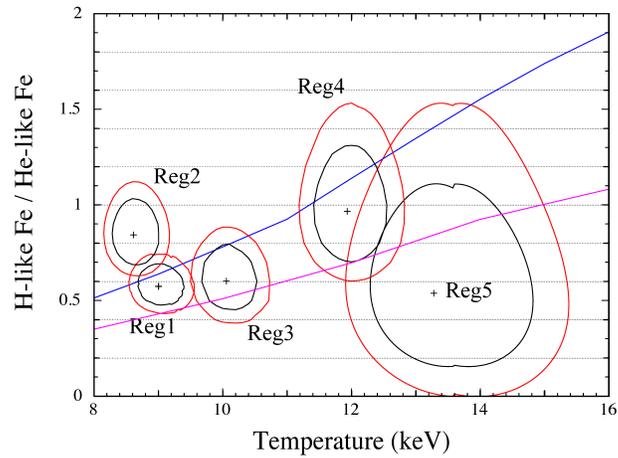} 
 \end{center}
\caption{Line ratios of each spectrum estimated by the empirical model (bremsstrahlung and gaussian components). The black and red regions represent the 68\% and 90\% confidence contours for the two interested parameters, respectively. The blue and magenta lines indicate the model curves in the case of the CIE state ($n_{\rm e}t=10^{13}$\,s\,cm$^{-3}$) and that of NEI state ($n_{\rm e}t=10^{12}$\,s\,cm$^{-3}$), respectively.}\label{fig:lineratio}
\end{figure}

Before detailed analyses by plasma emission models, we first examine the line intensity ratio of  Fe Ly\,$\alpha$ to Fe He\,$\alpha$ for the six regions. 
The ICM component is empirically modeled with a bremsstrahlung continuum and two Gaussian lines. 
The CXB component and the contamination from adjacent regions are considered as noted in section \ref{subsec:photon_contamination}. 
Figure \ref{fig:lineratio} shows the Fe Ly\,$\alpha$ / Fe He\,$\alpha$ line intensity ratio and the electron temperature of each region. 
The X-ray spectrum of Reg6 does not have enough statistics to constrain the line intensity ratio, and thus is not included in the plot. Respective blue and magenta curves indicate the line intensity ratio of Fe Ly\,$\alpha$ / Fe He\,$\alpha$ expected from the CIE model and that of the NEI model with $n_{\rm e}t=10^{12}$\,s\,cm$^{-3}$. We note that the line ratio in Reg5 shows a hint of NEI plasma, while the line ratio is almost consistent with that in the CIE state for other regions. We also note that the value of the Fe line ratio in Reg2 is significantly larger than the CIE model, which might imply an over-ionizing plasma.


\subsection{Fitting with a plasma emission model}
\label{subsec:NEIfit}

\begin{figure}
   \begin{minipage}{0.5\hsize}
   \begin{center}
      \includegraphics[width=8cm]{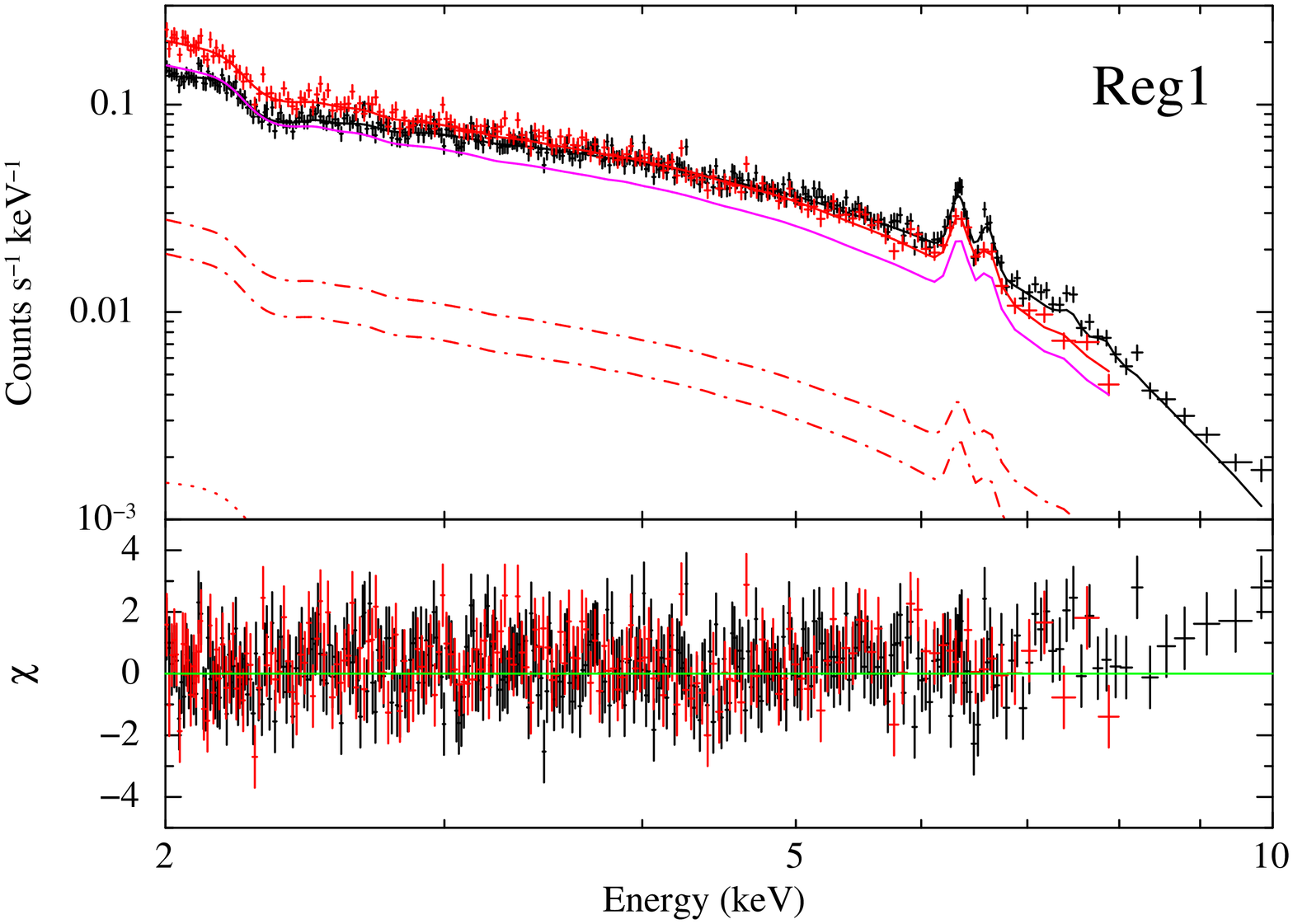}
   \end{center}
   \end{minipage}
      \begin{minipage}{0.5\hsize}
   \begin{center}
      \includegraphics[width=8cm]{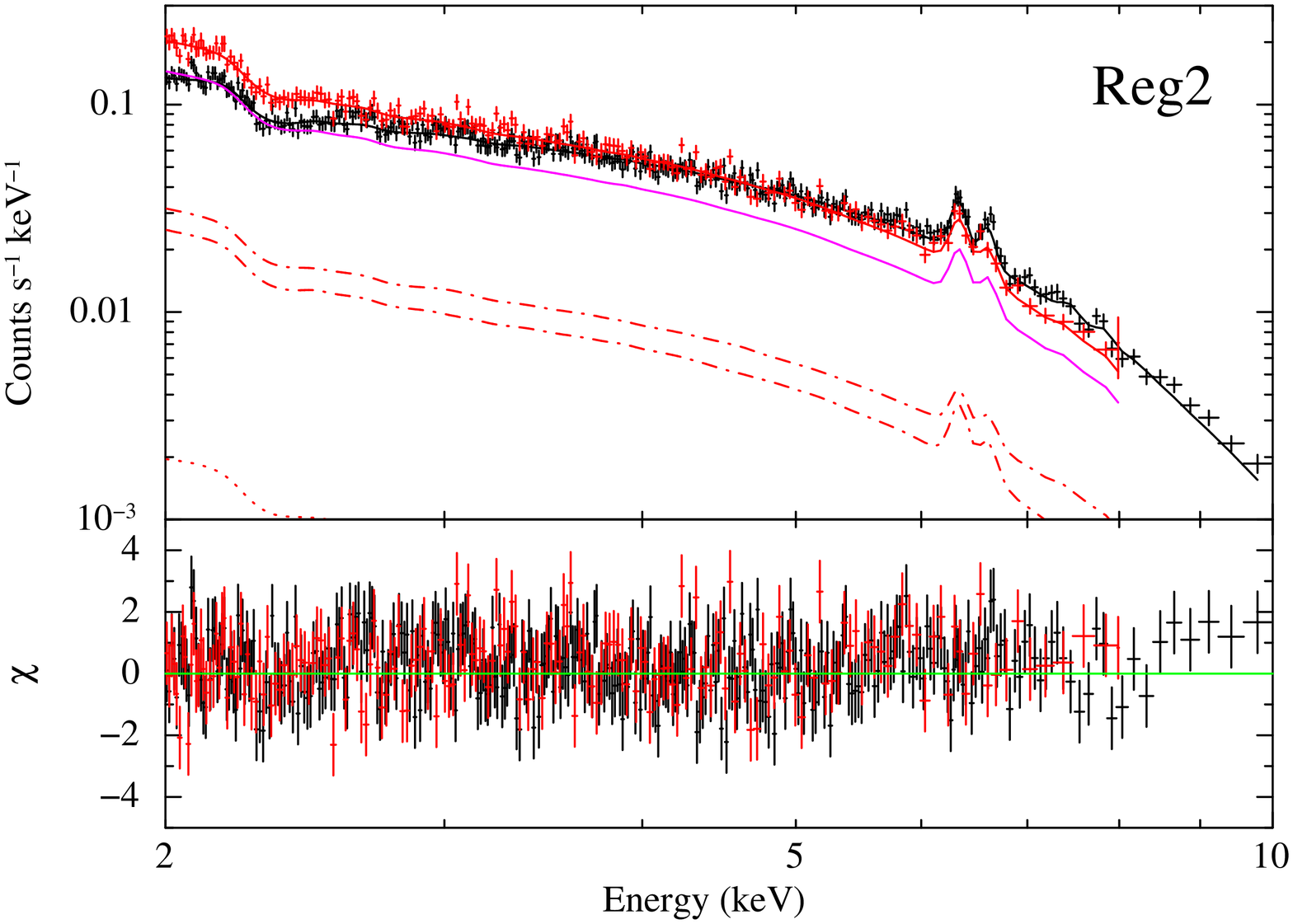}
   \end{center}
   \end{minipage}

   \begin{minipage}{0.5\hsize}
   \begin{center}
      \includegraphics[width=8cm]{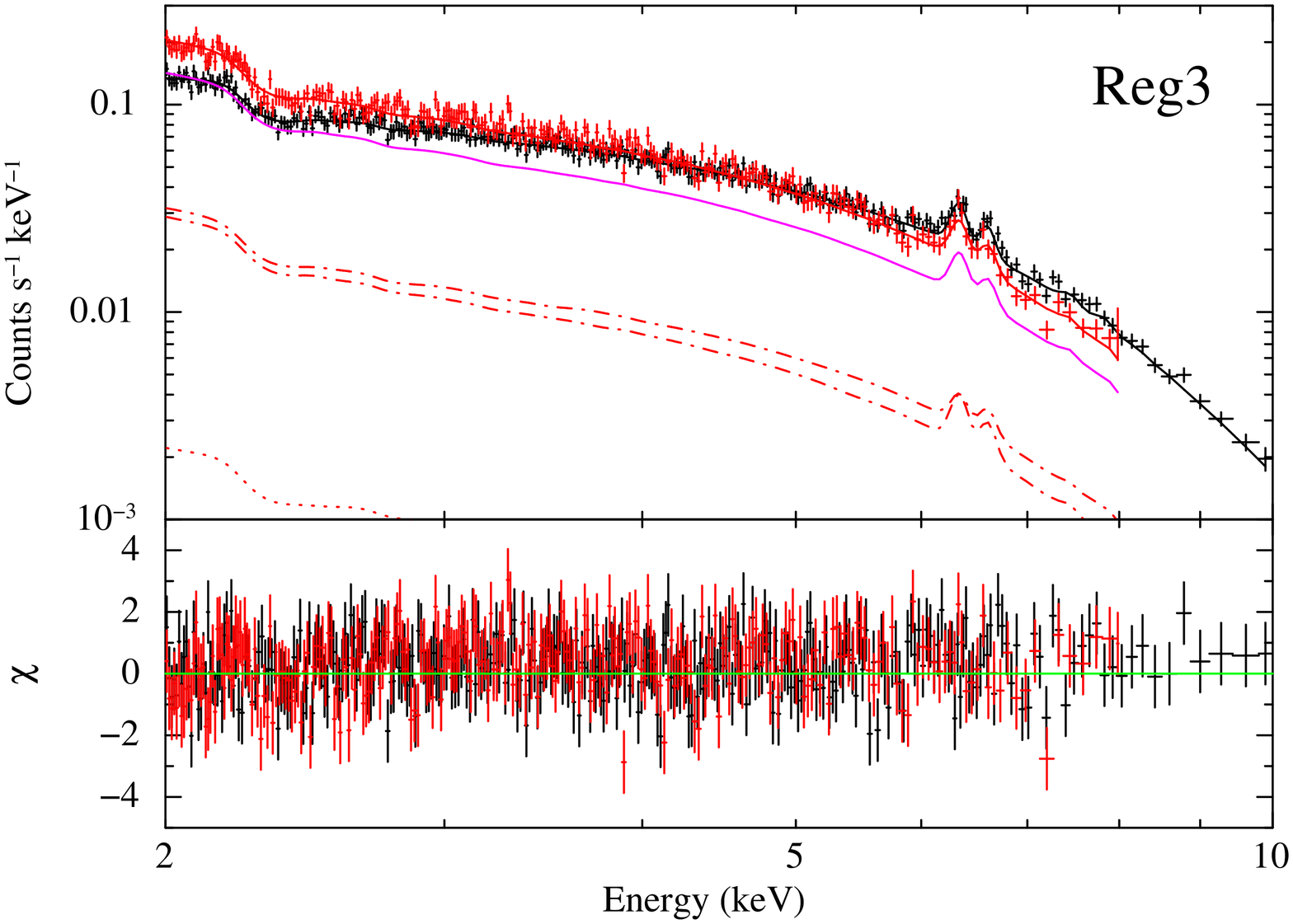}
   \end{center}
   \end{minipage}
      \begin{minipage}{0.5\hsize}
   \begin{center}
      \includegraphics[width=8cm]{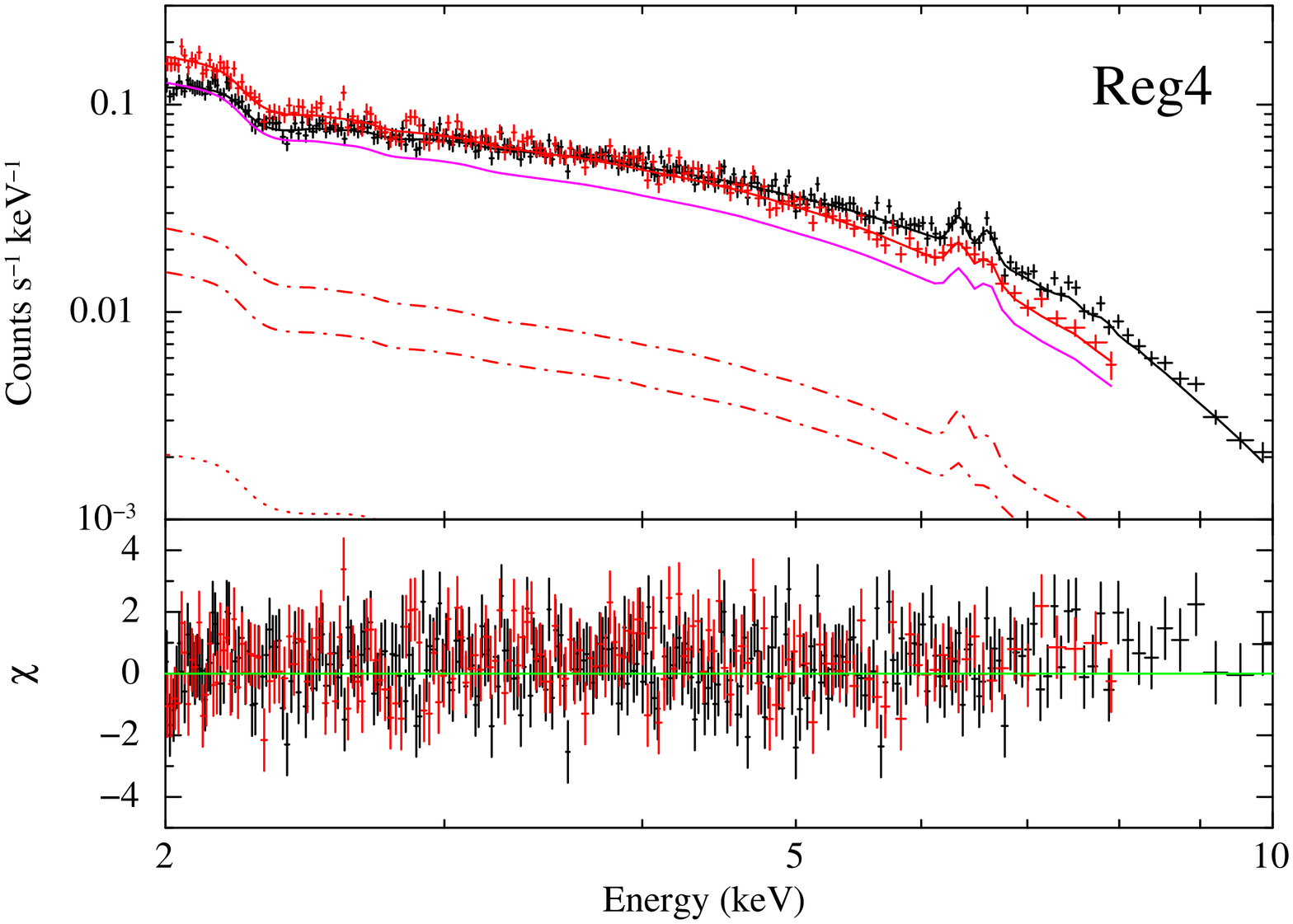}
   \end{center}
   \end{minipage}

   \begin{minipage}{0.5\hsize}
   \begin{center}
      \includegraphics[width=8cm]{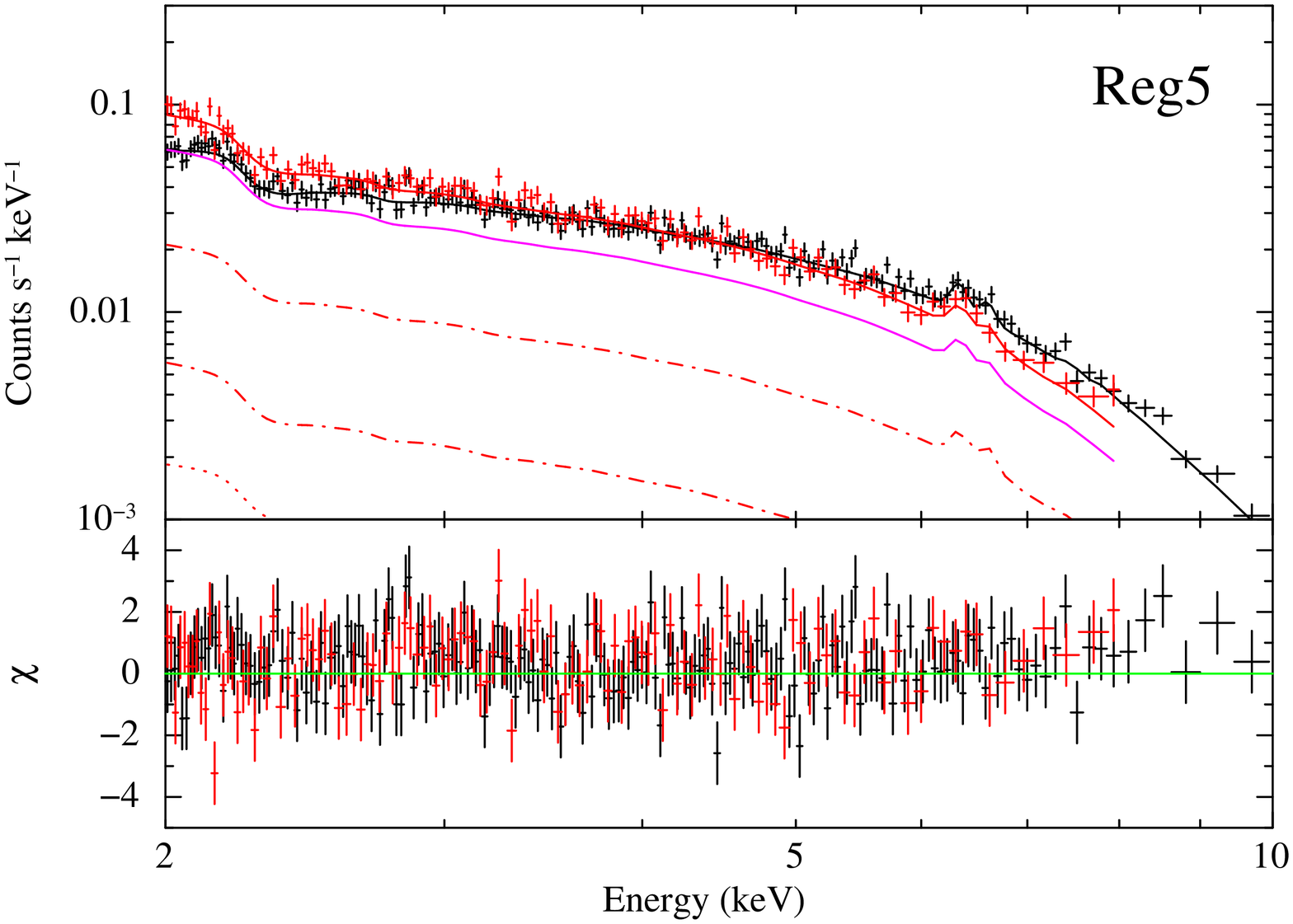}
   \end{center}
   \end{minipage}
      \begin{minipage}{0.5\hsize}
   \begin{center}
      \includegraphics[width=8cm]{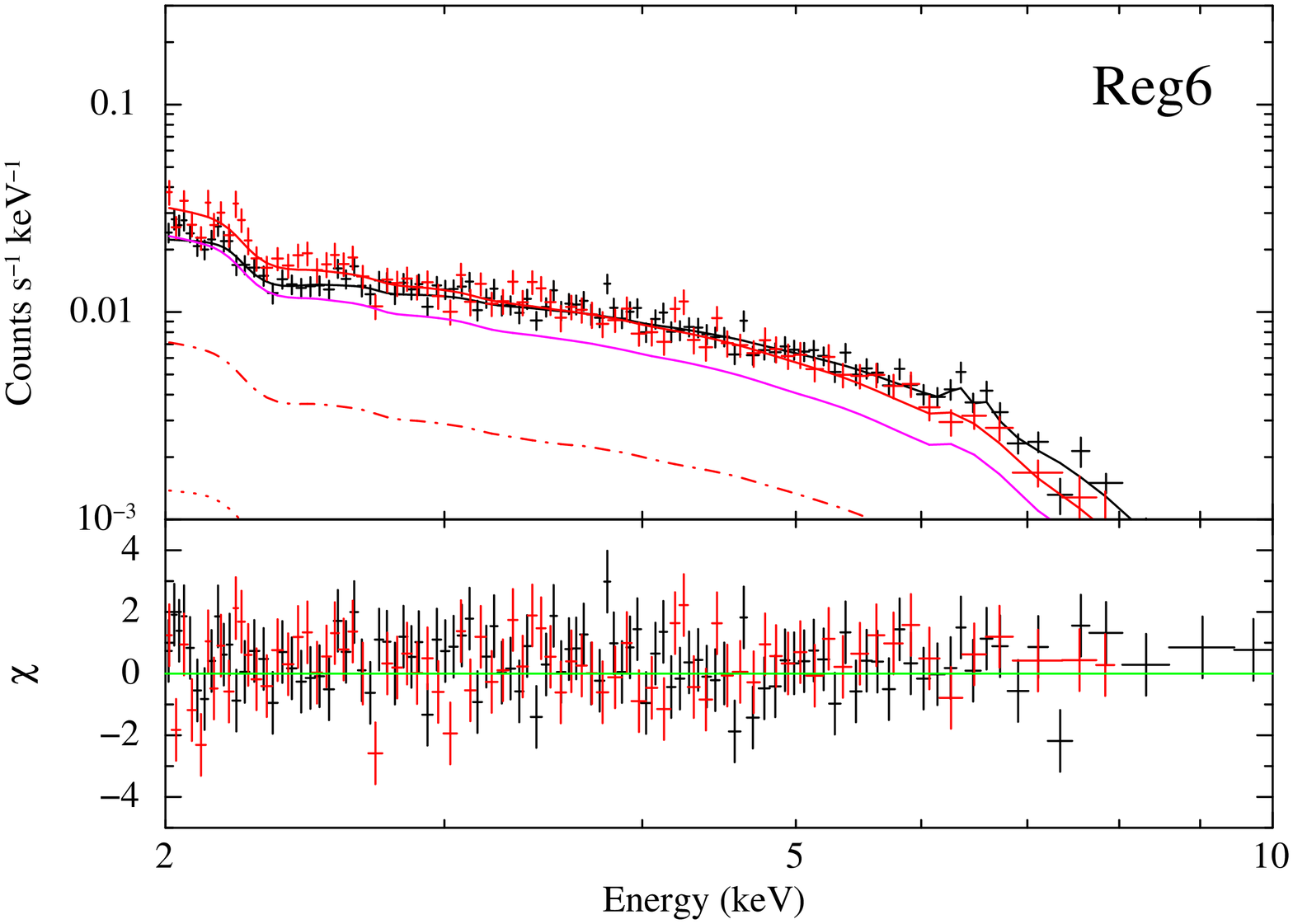}
   \end{center}
   \end{minipage}
 \caption{Spectra of each region we analyze in Abell 754. Each upper and lower panel show the spectrum with components of fitting model and the residual between data and the fitting model, respectively. The FI (black) and BI (red) spectra are fitted with the model consists of ICM (magenta solid line) + CXB (red dot line) + photon contaminations from adjacent regions (red dot-dash lines). Spectra are further rebinned for a display purpose.}
 \label{fig:A754spectra_XIS}
\end{figure}

We next perform individual spectral fitting by employing the XSPEC plasma code, NEI models for the ICM component. 
The redshift is fixed at the value ($z = 0.05393$) determined in the simultaneous fit described in section \ref{subsec:photon_contamination}. 
\citet{Henry_2004} and \citet{Macario_2011} show that the temperature in the outer (pre-shock) region is about 5\,keV (see region 12 in \cite{Henry_2004} and figure 3 in \cite{Macario_2011}). 
Thus, we assume the initial electron temperature of the NEI plasma to be 5\,keV. 
The X-ray spectrum of each region with its best fit model is shown in figure \ref{fig:A754spectra_XIS}, while the best fit parameters are summarized in table \ref{tab:NEIfit_indi}. 
Both the upper and lower bounds of the ionization parameter are constrained in Reg5, while only the lower limits are obtained in other regions. 

In this analysis, we fix the components of photon contamination from adjacent regions. 
However, since the photon contamination from Reg5 to Reg6 is significant to  the intrinsic Reg6 component, 
uncertainties in the parameters of the Reg6 component are large. 
Hence, we also fit the spectra of Reg5 and Reg6 simultaneously by setting the parameters of each emission model free, 
while the component of photon contamination from Reg4 to Reg5 are fixed. 
Even in this fitting the ionization parameters are constrained as shown in Table \ref{tab:NEIfit_Reg5_Reg6}, indicationg the NEI state. 

In the individual fitting of Reg5, we obtain the relatively low abundance comparing to Reg4 (Table \ref{tab:NEIfit_indi}).  
For reference, we perform the spectral fitting of Reg5 by fixing the abundance to the 90\% upper limits of abundance in Reg5, $0.168$ solar. 
In this case, the only lower limit of the ionization parameter is constrained ($10^{12.1} (>10^{11.8})$\,s\,cm$^{-3}$).

In these fitting, we fix the redshift to the value determined in the simultaneous fitting for 6 regions. 
If we set the redshift to be free for Reg5, the fitting result is $z=0.062^{+0.012}_{-0.009}$ and $n_{\rm e}t=10^{11.7} (>10^{11.0})$\,s\,cm$^{-3}$. 
Although this value of $n_{\rm e}t$ is consistent with the CIE, the redshifts of 0.062 is different from that for the common 6 regions by about 3000\,km/s. This difference is too large  for the cluster merging in the plane of the sky. 

\subsection{Spatial distribution of the ICM plasma parameters in Abell 754}
\label{subsubsec:spatial_distribution}

\begin{figure}
   \begin{minipage}{0.5\hsize}
   \begin{center}
      \includegraphics[width=8cm]{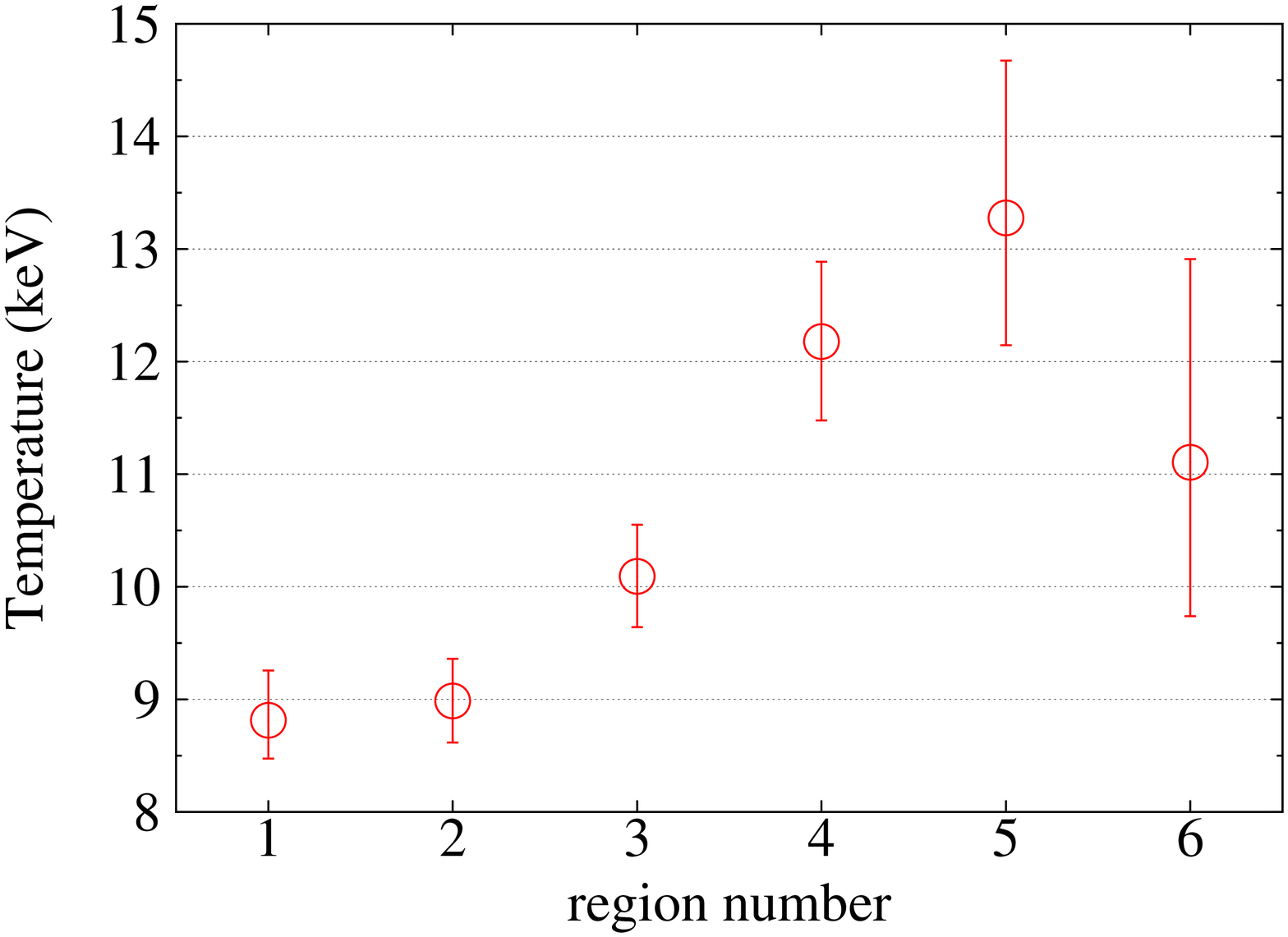}
   \end{center}
   \end{minipage}
      \begin{minipage}{0.5\hsize}
   \begin{center}
      \includegraphics[width=8cm]{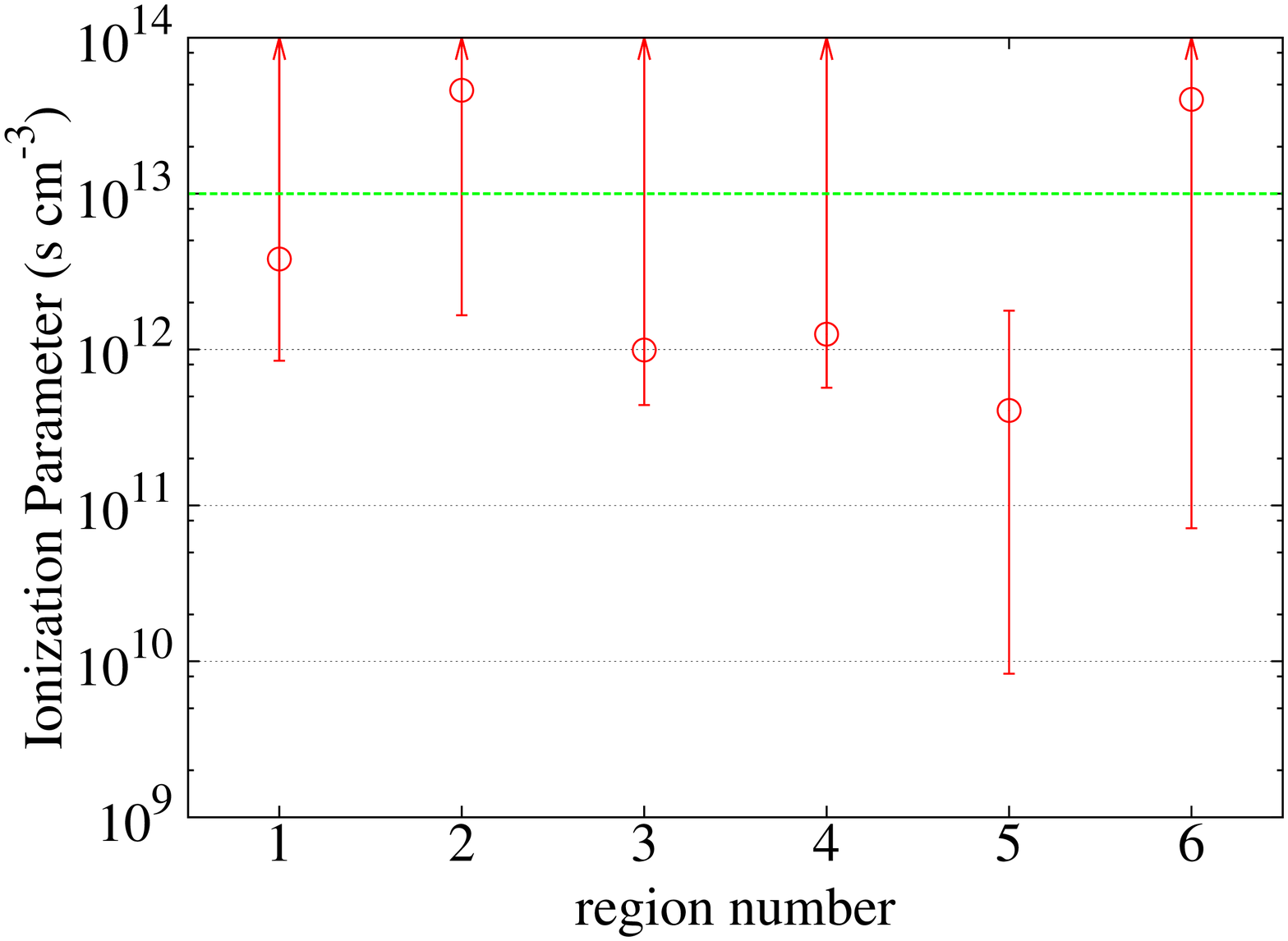}
   \end{center}
   \end{minipage}

   \begin{center}
      \includegraphics[width=8cm]{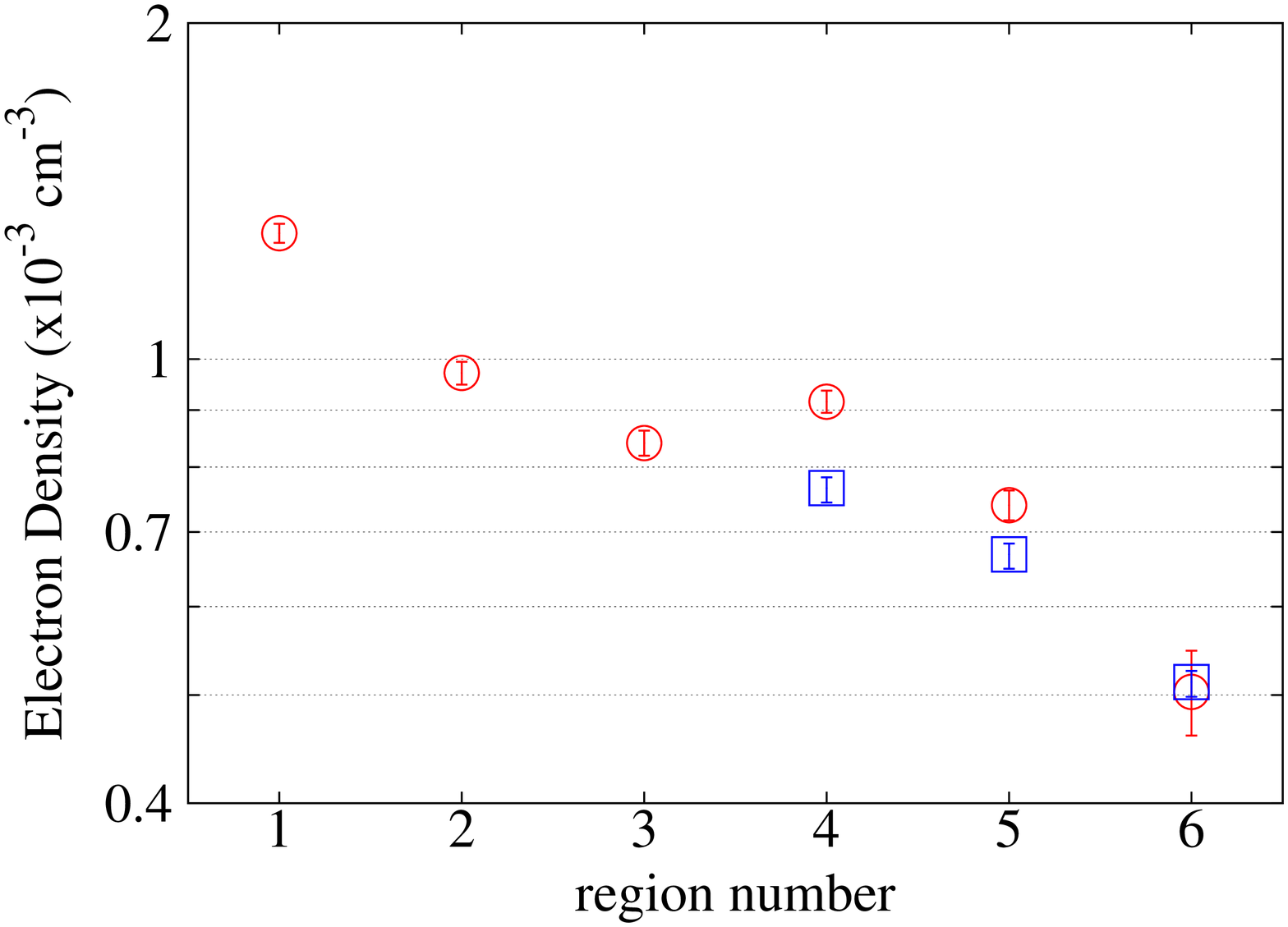}

   \end{center}

 \caption{Spatial distributions of the electron temperature (upper left), ionization parameter (upper right) and electron density (lower) estimated by model fittings in section \ref{subsec:NEIfit}. In each panel, error bars indicate the 90 \% confidence intervals. In the upper right panel, the green line indicates $n_{\rm e}t=10^{13}$\,s\,cm$^{-3}$. In the lower panel, the red and blue points represent the electron density with the cylindrical assumption and that with the spherical assumption, respectively.}
 \label{fig:spatial_distribution}
\end{figure}

We plot spatial distributions of the ICM plasma parameters determined in section \ref{subsec:NEIfit} and figure \ref{fig:spatial_distribution}. 
The upper left panel of figure \ref{fig:spatial_distribution} shows a temperature distribution. 
The temperature increases from Reg1 to Reg5, and the temperature of Reg6 is statistically equal to that of Reg5. 
The right panel of figure \ref{fig:spatial_distribution} shows the ionization parameters for each region. 
Only in Reg5, the upper bound of the ionization parameter is smaller than the CIE value (green line).

\begin{figure}
   \begin{center}
      \includegraphics[width=8cm]{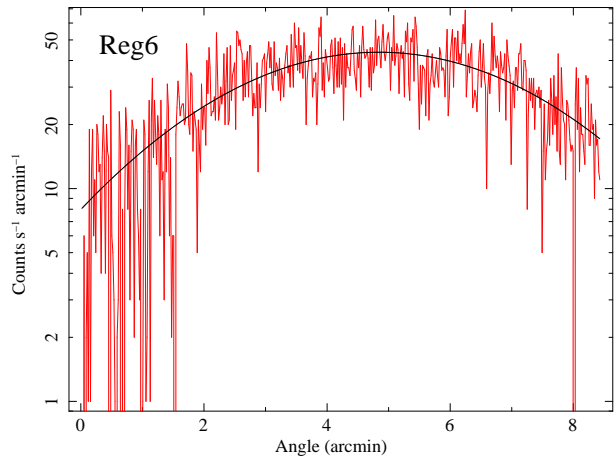}
   \end{center}
 \caption{Surface brightness distributions in Reg6 along the axis of red arrow in figure \ref{fig:A754Image}. Energy range is 2--8\,keV and the background is subtracted. The fitting model we applied (a single gaussian) is represented by the black solid line.}
 \label{fig:projected_surface_brightness}
\end{figure}

\begin{figure}
 \begin{center}
  \includegraphics[width=8cm]{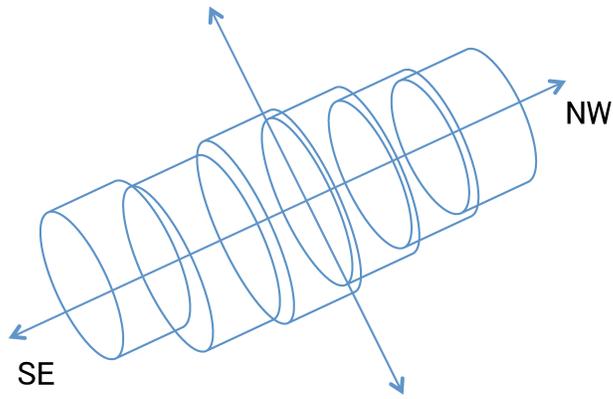}
 \end{center}
\caption{Geometry of the ICM in Abell 754 we assumed to estimate the electron density of each region. The radius of each cylinder is estimated by projected surface brightness (see section \ref{subsubsec:spatial_distribution}).}\label{fig:geometry}
\end{figure}

We next calculate the electron density of the ICM from the best fit values of the normalization in the NEI model. 
For this calculation, a three dimensional profile model of the ICM density is needed. 
This is not trivial, especially for merging clusters with complicated morphology as in our case. 
Figure \ref{fig:projected_surface_brightness} shows the background-subtracted one-dimensional surface brightness map of Reg6 projected to the direction perpendicular to the red axis in figure \ref{fig:A754Image}. 
For each of six region (Reg1--6), the surface brightness map is well-fitted by a Gaussian model. 
We assume that the ICM profile of Abell 754 can be approximated with a group of cylinders with different radii and axes as shown in figure \ref{fig:geometry} and each cylinder has a uniform density. 
The radius of each cylinder $R$ is calculated as $R=2\sigma_{\rm Gauss}$, where  $\sigma_{\rm Gauss}$ is the standard deviation of the Gaussian  model fit for the surface brightness, if we approximate the Gaussian profile with the hemisphere profile of the same variance. 
The best fit values of the radii and the volume of the cylinders are summarized in table \ref{tab:fit_result_cylinder}. 
The electron densities evaluated by using these cylindrical geometries are shown by red points in the bottom panel of figure \ref{fig:spatial_distribution} and in "cylindrical assumption" row of table \ref{tab:electron_density}. 
We found two peaks at Reg1 and Reg4 in the bottom panel of figure \ref{fig:spatial_distribution}. 
The significance of the second peak in Reg4 is $\sim7.9 \sigma$. 
The peak in Reg4 corresponds to the subcluster, which is consistent with the XMM-Newton observation (\cite{Henry_2004}). 

\begin{figure}
 \begin{center}
  \includegraphics[width=8cm]{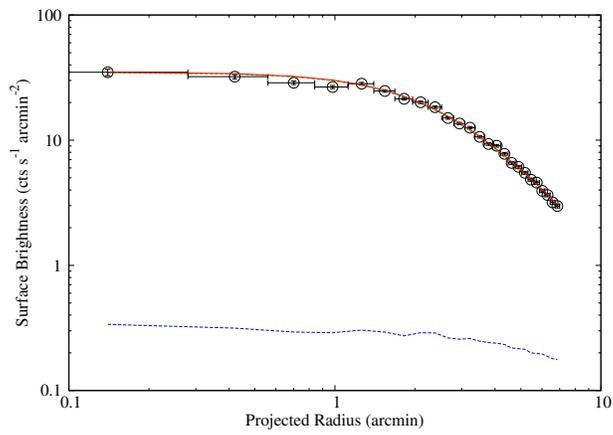}
 \end{center}
\caption{Fitting result of the surface brightness map in Reg4--Reg6 with the best fit $\beta$-model. The projected radius indicates the angular distance from the central position of Reg4 toward the Reg6 direction. Respective red and blue lines indicate the best fit model and the CXB component convolved by the Suzaku PSF.}\label{fig:beta_fit}
\end{figure}

In Reg4--Reg6, an alternative assumption of spherical symmetry could be applied. 
Assuming a $\beta$-model profile (\cite{Cavaliere_1978}) to a northwest part of the subcluster, we estimate the electron densities for Reg4--Reg6 as well. 
We use the $\beta$-model convolved by the Suzaku point spread function (PSF) to fit the radial profile as is done in \citet{Inoue_2014} and \citet{Mori_2013}. 
We estimate the Suzaku PSF with the FTOOL {\it xissimarfgen} (\cite{Ishisaki_2007}). 
Figure \ref{fig:beta_fit} shows the radial surface brightness map from the central position of Reg4 toward Reg6 with the best fit $\beta$-model of $r_c=2.67\pm0.01\,({\rm arcmin})=179.5\pm0.7\,({\rm kpc})$ and $\beta=0.444^{+0.002}_{-0.003}$. 
These parameters are consistent with those of a large sample of galaxy clusters (e.g., \cite{Mohr_1999}; \cite{Ota_2004}; \cite{Ota_2006}). 
Using this $\beta$-model, we average the electron density for each region within a radius of $3r_c$. 
The blue points in the bottom panel of figure \ref{fig:spatial_distribution} and "spherical assumption" row in table \ref{tab:electron_density} show the electron density estimated by the best fit $\beta$-model. 
The difference is only a factor of 1.0--1.2 between the electron densities under the spherical assumption and those under the cylindrical assumption. 
In the following discussion, we apply the values evaluated by the cylindrical assumption.

\subsection{Analysis of Reg5 with two temperatures CIE model}
\label{subsec:2CIEfit}

\begin{figure}
 \begin{center}
  \includegraphics[width=8cm]{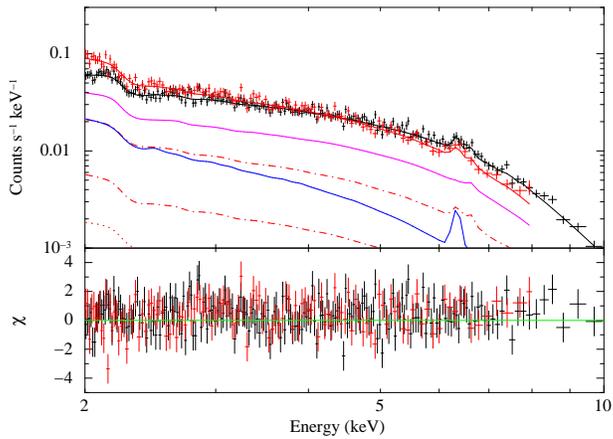} 
 \end{center}
\caption{Spectrum of Reg5 fitted with the two-temperature CIE model. Each mark or line is the same as those defined for figure \ref{fig:A754spectra_XIS}, except for that respective magenta and blue lines indicate the very high temperature and low temperature components, respectively.}\label{fig:twoCIEfit}
\end{figure}

We find an NEI plasma at Reg5 in section \ref{subsec:NEIfit}. Nevertheless, this result might depend on the ICM emission model employed. 
Although we assumed a single temperature NEI model in section \ref{subsec:NEIfit}, a two-temperature CIE plasma model might be another possibility. 
We then fit the X-ray spectrum of Reg5 with a combination of two CIE models for the ICM component. 
The abundances are linked between the two CIE models. 
Redshifts are commonly fixed to the value determined in section \ref{subsec:photon_contamination} ($z = 0.05393$). 
Figure \ref{fig:twoCIEfit} shows the spectrum of Reg5 with the best fit model under this assumption, while the best fit parameters are summarized in table \ref{tab:fit_result_etc}. 
The spectrum of Reg5 is well fitted with this two-temperature CIE model. 
Note, however, that the very high temperature ($kT>19.3$\,keV) component can be also fitted with a powerlaw model.

\subsection{Analysis of the HXD data}
\label{subsec:HXDfit}

As mentioned in section \ref{sec:abell754}, Abell 754 hosts a radio halo extended over almost its whole region (\cite{Macario_2011}). 
It means there exist high-energy electrons, which can reproduce a non-thermal hard X-ray emission via an inverse Compton process. 
The HXD on Suzaku, which simultaneously observes the same source as the XIS, is useful to study this component. 
The dead-time corrected raw HXD count rate in 15--40\,keV energy band is $0.3586\pm0.0019$\,counts\,s$^{-1}$, where the 1 $\sigma$ statistical error is shown, while the NXB and the CXB modeled count rate are $0.2998\pm0.0006$\,counts\,s$^{-1}$ and $0.0161$\,counts\,s$^{-1}$ (\cite{Boldt_1987}), respectively. 
The net count rate in which the NXB and the CXB are subtracted is $0.04287\pm0.00206$\,counts\,s$^{-1}$, corresponding to 14.3\% of NXB. 
This is significantly larger than systematic errors in the NXB model (\cite{Fukazawa_2009}), indicating that Abell 754 is positively detected with the HXD.

Our next concern is whether this hard X-ray emission can be reproduced only with an extension of the thermal components observed with the XIS or an extra non-thermal component is required. 
We consider a spectral model consisting of the CXB and seven thermal components (Reg0--6) determined with the XIS spectra as described in section \ref{subsec:NEIfit}. 
The parameters of all the thermal components and those of CXB are fixed to the values shown in table \ref{tab:NEIfit_indi} and the values reported by \citet{Boldt_1987}, respectively, allowing only one free parameter to tune the entire normalization between the HXD and the XIS. 
We prepare the HXD ARFs for each spatial region by using {\it hxdarfgen} and apply them to the corresponding thermal component.

\begin{figure}
 \begin{center}
  \includegraphics[width=8cm]{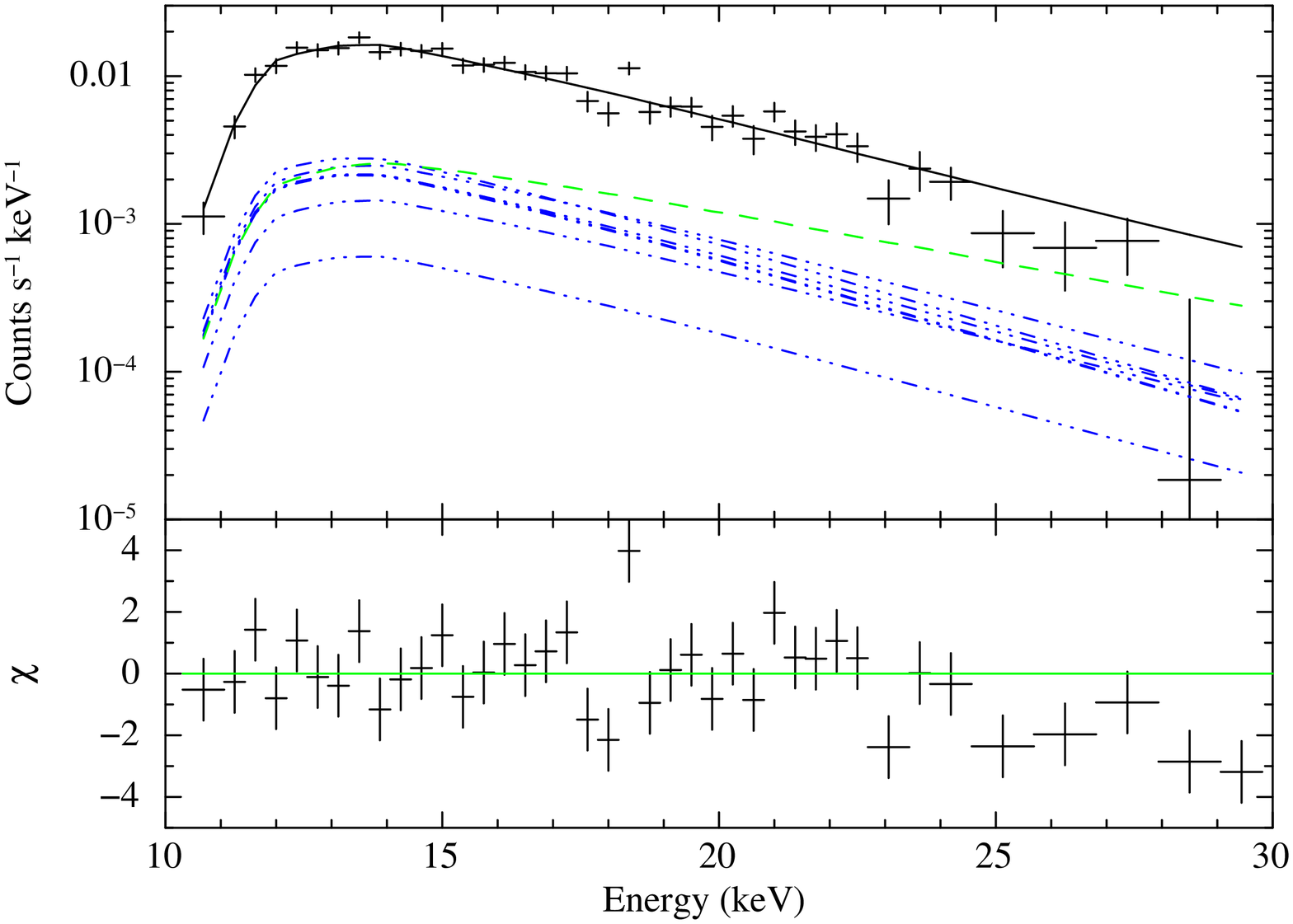} 
 \end{center}
\caption{HXD spectrum of Abell 754. Upper panel and lower panel show the spectrum with individual components of fitting model (black line) and the residual between data and the fitting model, respectively. Green dash line and seven dot-dot-dot-dash lines represent the component of the CXB and the thermal components of each region (Reg0--6), repectively. }\label{fig:HXDspectrum}
\end{figure}

Figure \ref{fig:HXDspectrum} shows the fitting result in the 10--30\,keV band. 
The best fit values of the normalization factor for the HXD is $0.969_{-0.033}^{+0.033}$ ($\chi^2$/d.o.f.$=85.91 / 51$). 
According to the Suzaku memo\footnote{http://www.astro.isas.jaxa.jp/suzaku/doc/suzakumemo/suzakumemo-2007-11.pdf}, the systematic difference between the HXD and the XIS is $1.132\pm0.014$, which is the expected value of the normalization factor we introduced in the fitting. 
The systematic uncertainty of the NXB model of $2.1-2.7$\% (\cite{Fukazawa_2009}) yields systematic error of 0.2 for the normalization. 
The difference between observed and expected values for the normalization is smaller than this value. 
We thus conclude that the HXD spectrum is reproduced only with the thermal components we modeled for the XIS spectra. 
The hard X-ray excess ($>45$\,keV) such as reported by \citet{Fusco-Femiano_2003} is not needed to reproduce the HXD spectrum in the 10--30\,keV band.

We also note that the counting statistic of our HXD data does not allow us to distinguish two possibilities, i.e., the single temperature NEI model or the two-temperature CIE model wit respective to Reg5.

\section{Discussion}

We have performed the spectral analysis of Abell 754 with the NEI plasma model. 
The spatial distribution of the electron density shows two peaks at Reg1 and Reg4. 
The peak at Reg1 corresponds to the main cluster, while that at Reg4 does the subcluster, both of which are merging. 
This is consistent with the previous view suggested by the XMM-Newton observation of this cluster (\cite{Henry_2004}). 
The temperature distribution of Abell 754 shows a gradient along the merging direction and takes its highest value of $13.3_{-1.1}^{+1.4}$\,keV at Reg5.

The main finding of this paper is the NEI plasma at Reg5. 
This is suggested both from the line intensity ratio between Fe He\,$\alpha$ and Fe Ly\,$\alpha$ in the empirical model fitting and from the ionization parameter $n_{\rm e}t$ derived in the single temperature NEI plasma model fitting. 
We, however, cannot rule out the possibility that the ICM component in Reg5 consists of two CIE components with different temperatures.

If the single temperature NEI model is adopted, we are able to discuss past history of the plasma in Reg5 quantitatively as we expected. 
With the ionization parameters we obtained ($n_{\rm e}t=10^{11.6_{-1.7}^{+0.6}}$\,s\,cm$^{-3}$) and the electron density we evaluated ($n_{\rm e}=0.740\pm0.023 \times 10^{-3}$\,cm$^{-3}$), the timescale after the plasma heating is obtained to be 0.36--76 Myr at the 90\% confidence level. 
This is rather short comparing with the evolutional timescale of clusters. 
However, if there is another shock wave at the boundary between Reg5 and Reg6 (we call it ''shock-NW''), in addition to shock-SE identified by \citet{Macario_2011} near Reg0, this timescale is consistent with the time after the shock passage. 
Although the errors in the temperatures derived in the fit and the spatial resolution of Suzaku are not small enough to constrain, a possible temperature jump from 14\,keV at Reg5 to 10\,keV at Reg6 is not ruled out. 
If there is the temperature jump, we can apply the Rankine-Hugoniot condition with an adiabatic index $\gamma=5/3$. 
This temperature jump corresponds to the shock with the Mach number relative to the pre-shock region $\sim1.4$ and that relevant to the post-shock region $\sim0.89$. 
Since the post-shock velocity is ~1300\,km/s, the timescale that shock-NW passes through the half of Reg5 is $\sim$69 Myr, consistent with the timescale we derived from the ionization parameter.

In the spectral analysis with plasma models, we have roughly assumed the initial temperature is 5\,keV. 
Initial temperatures slightly alter the ionization timescale. 
We fit the spectra using the plasma model with different initial temperatures, 0.1, 1, 7\,keV. 
As shown in table \ref{tab:fit_result_kTinit}, the timescales in almost cases are consistent with that for shock-NW passing through the region. 

The temperature jump between Reg5 and Reg6 is just our speculation, and we have to wait future observations. 
Nevertheless, \citet{Macario_2011} suggested a coincidence between the edge of the radio emission and shock-SE they identified with the Chandra data near Reg0. 
We notice that the other edge of the radio emission roughly corresponds to shock-NW, indirectly supporting our hypothesis.

We also notice a hint of over-ionized plasma at Reg2 in figure \ref{fig:lineratio}. 
Applying an over-ionized plasma model with an initial temperature of $kT_{\rm init}=15$\,keV to the ICM component of this spectrum, we obtain the relaxation timescale of $n_{\rm e}t=10^{11.78^{+0.40}_{-0.34}}$\,s\,cm$^{-3}$. 
While it is difficult to explain a mechanism to make an over-ionized plasma in the ICM, we notice that the simulation of merging clusters by \citet{Akahori_Yoshikawa_2010} shows specific regions where the Fe line ratio is consistent with the over-ionized plasma (see figure 6 supplied by \cite{Akahori_Yoshikawa_2010}). 
This feature is more prominent in the case of the collision between clusters with different masses. 
The possibility of an over-ionized plasma in the ICM is an issue for future study.

Go back to the ICM model at Reg5, another interpretation is a two-temperature CIE component. 
The best fit values of the temperatures are $4.1_{-1.5}^{+4.2}$\,keV and $31.5 (>19.3)$\,keV. 
Although the averaged temperature of entire Abell 754 is $\sim 9$\,keV, we cannot deny the possibility that the cool component is a remnant of a cool core before merging. 
In fact, setting the initial condition of the temperatures to $\sim 3$\,keV and $\sim 7$\,keV, \citet{Roettiger_1998} well reproduced the temperature map of Abell 754 by numerical simulation. 
The component of the very high temperature 
suggests there may be extremely hot region at Reg5 which is strongly heated by a merger (see \cite{Ota_2008}). 
Such high temperature component is also approximated with a powerlaw model as reported by \citet{Fusco-Femiano_2003}, although it is not detected significantly in the HXD spectrum in the 10--30\,keV energy band (section \ref{subsec:HXDfit}). 
The Hard X-ray imaging performed by NuSTAR (\cite{Harrison_2013}) or hard X-ray imager installed on ASTRO-H (\cite{Takahashi_2012}) will help us to distinguish these hypotheses. 

Whichever interpretation, the single temperature NEI or the two temperature CIE, is correct, we consider the merging process cause a deviation from the single temperature CIE model at Reg5. 
Discrimination of these two with future observations must be important to understand the merging history and physics in this cluster, and clusters in general.

\section{Summary}

We performed the X-ray spectral analysis of Abell 754 observed with Suzaku. 
Assuming a single temperature plasma model at each region, we measured the spatial distributions of plasma parameters along the merger axis. 
The temperature increases from southeast to northwest and highest at Reg5 ($kT=13.3^{+1.4}_{-1.1}$\,keV). 
The ionization parameter deviates from its equilibrium value only at Reg5 ($n_{\rm e}t=10^{11.6_{-1.7}^{+0.6}}$\,s\,cm$^{-3}$), suggesting the NEI plasma at that region. 
The timescale after the plasma heating in this region is estimated from the ionization parameter, and is 0.36--76 Myr at the 90\% confidence level. 
This timescale is rather short but consistent with the time after the shock passage, if we assume the shock front at the boundary of Reg5 and Reg6 (shock-NW). 
On the other hand, another possibility in which the ICM at Reg5 consists of two-temperature CIE plasma is not ruled out. 
Even in that case, we suspect the very high  ($>19.3$\,keV) temperature component, which might also be a non-thermal component, is due to merging process in this cluster.

\begin{ack}
We thank all members of the Suzaku operation and calibration teams. This work is sup-
ported by Japan Society for the Promotion of Science (JSPS) KAKENHI Grant Number 
15K17610, 23000004, 23340071, 24540229, 26109506
\end{ack}





\newpage

\begin{table}
  \tbl{Result of individual fittings for six regions of Abell 754 with the NEI model.}{%
  \begin{tabular}{lllllll}
      \hline
      &Reg1&Reg2&Reg3&Reg4&Reg5&Reg6 \\ \hline \hline
      $kT$ (keV)&$8.81_{-0.34}^{+0.44}$&$8.99\pm0.37$&$10.09_{-0.45}^{+0.46}$&$12.18_{-0.70}^{+0.71}$&$13.3_{-1.1}^{+1.4}$&$11.1_{-1.4}^{+1.8}$ \\
      $kT_{\rm init}$ (keV)&\multicolumn{6}{c}{$5.0$ (fix)} \\
      abundance (solar)&$0.294_{-0.050}^{+0.030}$&$0.239_{-0.027}^{+0.029}$&$0.177_{-0.041}^{+0.053}$&$0.178_{-0.051}^{+0.062}$&$0.098_{-0.048}^{+0.070}$&$0.129_{-0.085}^{+0.087}$ \\
      $n_{\rm e}t$ ($10^{12}$\,s\,cm$^{-3}$)&$3.81(>0.85)$&$46.02(>1.66)$&$0.99(>0.44)$&$1.25(>0.57)$&$0.41_{-0.40}^{+1.37}$&$40.20(>0.07)$ \\
      redshift&\multicolumn{6}{c}{$0.05393$ (fix)} \\
      norm\footnotemark[$\dagger$] ($10^{-2}$)&$1.398\pm0.019$&$1.059\pm0.015$&$0.911\pm0.013$&$0.871_{-0.012}^{+0.011}$&$0.470\pm0.008$&$0.228_{-0.007}^{+0.008}$ \\ \hline
      $\chi^2$ / d.o.f.&2308.26 / 2334&2394.41 / 2389&2325.27 / 2443&2342.50 / 2357&1578.95 / 1592&683.84 / 719 \\ \hline
    \end{tabular}}\label{tab:NEIfit_indi}
\begin{tabnote}
      \footnotemark[$\dagger$]: ${\rm norm}=\int n_{\rm e}n_{\rm H}dV/(4\pi(D_A(1+z))^2)\times 10^{-14}$ cm$^{-5}$, where $D_A$\,(cm) is the angular diameter distance to the source.
\end{tabnote}
\end{table}

\begin{table}
  \tbl{Result of simultaneous fitting between Reg5 and Reg6 spectra with the NEI model.}{%
  \begin{tabular}{lll}
      \hline
      &Reg5&Reg6 \\ \hline \hline
      $kT$ (keV)&$13.5_{-1.2}^{+1.5}$&$11.4_{-1.4}^{+1.8}$ \\
      $kT_{\rm init}$ (keV)&\multicolumn{2}{c}{$5.0$ (fix)} \\
      abundance (solar)&$0.098_{-0.048}^{+0.069}$&$0.129_{-0.086}^{+0.088}$ \\
      $n_{\rm e}t$ ($10^{12}$\,s\,cm$^{-3}$)&$0.41_{-0.38}^{+1.24}$&$40.20(>0.07)$ \\
      redshift&\multicolumn{2}{c}{$0.05393$ (fix)} \\
      norm\footnotemark[$\dagger$] ($10^{-2}$)&$0.471\pm0.008$&$0.243\pm0.006$ \\ \hline
      $\chi^2$ / d.o.f.&\multicolumn{2}{c}{2264.12 / 2312} \\ \hline
    \end{tabular}}\label{tab:NEIfit_Reg5_Reg6}
\begin{tabnote}
      \footnotemark[$\dagger$]: ${\rm norm}=\int n_{\rm e}n_{\rm H}dV/(4\pi(D_A(1+z))^2)\times 10^{-14}$ cm$^{-5}$, where $D_A$\,(cm) is the angular diameter distance to the source.
\end{tabnote}
\end{table}

\begin{table}
  \tbl{The fitting result of cylinder we assume as Abell 754 geometry.}{%
  \begin{tabular}{lllllll}
      \hline
      & Reg1 & Reg2 & Reg3 & Reg4 & Reg5 & Reg6 \\ \hline \hline
      R (arcmin)&4.952$^{+0.050}_{-0.048}$&5.799$^{+0.069}_{-0.067}$&6.243$^{+0.081}_{-0.079}$&5.597$^{+0.065}_{-0.063}$&5.128$^{+0.080}_{-0.078}$&5.25$^{+0.24}_{-0.22}$ \\
      Volume ($\times10^{72}\,$cm$^3$)&5.81$^{+0.12}_{-0.11}$&7.97$^{+0.19}_{-0.18}$&9.23$^{+0.24}_{-0.23}$&7.42$\pm0.17$&6.23$^{+0.20}_{-0.19}$&6.52$^{+0.61}_{-0.53}$ \\ \hline
    \end{tabular}}\label{tab:fit_result_cylinder}
\begin{tabnote}
\end{tabnote}
\end{table}

\begin{table}
  \tbl{Electron density $n_{\rm e}$ of each region with different assumptions.}{%
  \begin{tabular}{lllllll}
      \hline
      & Reg1 & Reg2 & Reg3 & Reg4 & Reg5 & Reg6 \\ \hline \hline
      \multicolumn{7}{c}{cylindrical assumption} \\ 
      $n_{\rm e}$ (10$^{-3}$\,cm$^{-3}$) & 1.296$\pm0.026$ & 0.972$\pm0.023$ & 0.841$^{+0.022}_{-0.021}$ & 0.916$\pm0.021$ & 0.740$\pm0.023$ & 0.503$^{+0.045}_{-0.043}$ \\ \hline
      \multicolumn{7}{c}{spherical assumption} \\
      $n_{\rm e}$ (10$^{-3}$\,cm$^{-3}$) & \multicolumn{1}{c}{---} & \multicolumn{1}{c}{---} & \multicolumn{1}{c}{---} & $0.766^{+0.017}_{-0.022}$ & $0.668^{+0.015}_{-0.019}$ & $0.514^{+0.012}_{-0.015}$ \\ \hline
    \end{tabular}}\label{tab:electron_density}
\begin{tabnote}
\end{tabnote}
\end{table}


\begin{table}
  \tbl{Fitting results of Reg5 spectrum with two-temperature CIE model.}{%
  \begin{tabular}{lll}
      \hline
      & \multicolumn{2}{c}{Reg5}  \\ \hline \hline
      $kT$ (keV) & $4.1_{-1.5}^{+4.2}$ & $31.5 (>19.3)$ \\
      abundance (solar) & \multicolumn{2}{c}{$0.203_{-0.081}^{+0.124}$} \\
      redshift & \multicolumn{2}{c}{$0.05393$ (fix)} \\
      norm\footnotemark[$\dagger$] ($10^{-2}$) & $0.175_{-0.072}^{+0.113}$ & $0.344_{-0.094}^{+0.050}$ \\ \hline
      $\chi^2$ / d.o.f. & \multicolumn{2}{c}{1575.25 / 1591} \\ \hline
    \end{tabular}}\label{tab:fit_result_etc}
\begin{tabnote}
      \footnotemark[$\dagger$]: ${\rm norm}=\int n_{\rm e}n_{\rm H}dV/(4\pi(D_A(1+z))^2)\times 10^{-14}$ cm$^{-5}$, where $D_A$\,(cm) is the angular diameter distance to the source.
\end{tabnote}
\end{table}

\begin{table}
  \tbl{Fitting results with different assumptions of the initial temperature.}{%
  \begin{tabular}{lllll}
      \hline
      $kT_{\rm init}$ & 0.1\,keV & 1\,keV & 5\,keV & 7\,keV  \\ \hline \hline
      $n_{\rm e}t$ ($10^{12}$\,s\,cm$^{-3}$) & $0.70_{-0.39}^{+1.45}$ & $0.66_{-0.39}^{+1.44}$ & $0.41_{-0.40}^{+1.37}$ & $0.19 (<1.50)$ \\
      timescale\footnotemark[$\dagger$] (Myr) & 13.1--92.2 & 11.6--90.0 & 0.36--76 & 0--64.3 \\ \hline
    \end{tabular}}\label{tab:fit_result_kTinit}
\begin{tabnote}
      \footnotemark[$\dagger$]: We express the timescales from the shock heating as 90\% confidence levels.
\end{tabnote}
\end{table}



\end{document}